\documentclass{article}

\usepackage[utf8]{inputenc}
\usepackage{graphicx,graphics}
\usepackage{indentfirst}
\usepackage{caption}
\usepackage{subfig}
\usepackage{cite}
\usepackage{amssymb, url}
\usepackage{comment}
\usepackage{lscape}
\usepackage{longtable}
\usepackage{array}
\usepackage[title]{appendix}
\usepackage{amsmath,amsfonts}
\usepackage{setspace}
\usepackage[subfigure]{tocloft}
\usepackage{lineno}
\usepackage{authblk}
\usepackage[top=3cm, bottom=3cm, left=2.5cm, right=2.5cm]{geometry}
\usepackage{setspace}
\begin{document} 
\title{Signal-carrying speckle in Optical Coherence Tomography: a methodological review on biomedical applications}
\date{}

\author[a,b]{Vania Bastos Silva}
\author[b,*]{Danilo Andrade De Jesus}
\author[b]{Stefan Klein}
\author[b]{Theo van Walsum}
\author[a]{João Cardoso}
\author[b]{Luisa Sánchez Brea}
\author[a]{Pedro G. Vaz}

\affil[a]{Laboratory for Instrumentation, Biomedical Engineering and Radiation Physics (LIBPhys-UC), Department of Physics, University of Coimbra, Coimbra, Portugal}
\affil[b]{Biomedical Imaging Group Rotterdam, Department of Radiology \& Nuclear Medicine, Erasmus MC, University Medical Center Rotterdam, the Netherlands}
\affil[*]{d.andradedejesus@erasmusmc.nl}

\maketitle

\begin{abstract}

Significance: Speckle has historically been considered a source of noise in coherent light imaging. However, a number of works in optical coherence tomography (OCT) imaging have shown that speckle patterns may contain relevant information regarding sub-resolution and structural properties of the tissues from which it is originated.

Aim: The objective of this work is to provide a comprehensive overview of the methods developed for retrieving speckle information in biomedical OCT applications.

Approach: PubMed and Scopus databases were used to perform a systematic review on studies published until April 2021. From 134-screened studies, 37 were eligible for this review.

Results: The studies have been clustered according to the nature of their analysis, namely static or dynamic, and all features were described and analysed. The results show that features retrieved from speckle can be used successfully in different applications, such as classification and segmentation. However, the results also show that speckle analysis is highly application-dependant, and the best approach varies between applications. 

Conclusions: Several of the reviewed analysis were only performed in a theoretical context or using phantoms, showing that signal-carrying speckle analysis in OCT imaging is still in its early stage, and further work is needed to validate its applicability and reproducibility in a clinical context.

\end{abstract}


\section{Introduction}\label{sec:introduction}
Optical Coherence Tomography (OCT) is an optical imaging modality based on low-coherence interferometry. It is a non-invasive technique that provides \textit{in vivo} cross sectional images of microscopic structures with high spatial and temporal resolutions, making it an appealing technique for multiple areas in pre-clinical and clinical research \cite{seevaratnam2014quantifying}.

\begin{figure}[t]
\centering
\includegraphics[width=0.45\textwidth]{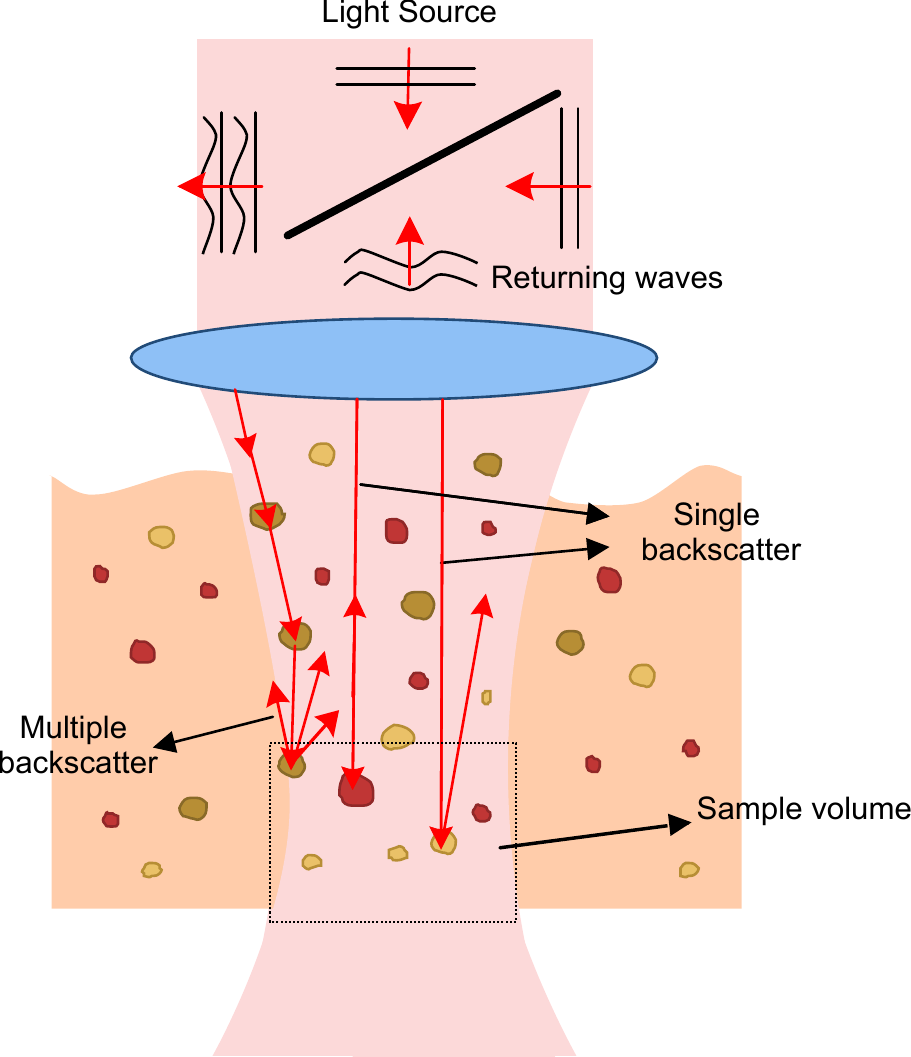}
\caption{Process of speckle formation in OCT. Speckle patterns result from coherent superposition of multiple backscattered waves from particles in the sample volume.}
\label{fig:speckleformation}
\end{figure}

In OCT imaging, the tissue is scanned by an optical beam, and most of the light is either refracted or scattered. The incident light travels through different optical paths, with different lengths, until it reaches the image plane. The light intensity at each point of the plane results from destructive/constructive interference of all light waves at that single point. This phenomenon, illustrated in Figure \ref{fig:speckleformation}, creates granular patterns, known as speckle patterns. Speckle appears everywhere when an optically rough surface is illuminated with coherent light, making it common to all coherent imaging modalities \cite{gossage2003texture}.

Schmitt \textit{et al.} \cite{schmitt1999speckle} were among the first to discuss the OCT speckle origin, distinguishing between two types of speckle that appear on a cross-sectional image: multiple backscattering of the light beam, and delays caused by multiple forward scattering. Since then, a number of works have interpreted speckle as a source of noise in OCT imaging, as it reduces the image quality and contrast, making boundaries between tissues less distinguishable. Because of this, methods to suppress and reduce speckle have been developed, including filtering \cite{ozcan2007speckle}, averaging \cite{szkulmowski2012efficient} or wavelet processing techniques \cite{xiang1998speckle}.

Speckle patterns change depending on different parameters, such as the properties of the light source, the propagating beam, the aperture of the detector, and the inner properties and structural organization of the tissues \cite{jesus2017assessment, vaz2016laser}. The latter indicates that speckle may contain relevant information regarding sub-resolution and structural properties of the tissues from which it originated \cite{de2015new}. In fact, in the work presented by Schmitt \textit{et al.} \cite{schmitt1999speckle}, speckle patterns in OCT are already mentioned as having a dual role, both as a source of noise, signal-degrading speckle, and as a carrier of information, signal-carrying speckle. This indicates that, besides the granular noise observed in OCT raw images, the imaged speckle also carries information. This information may be used to characterize  the imaged tissue \cite{kasaragod2010speckle}. 

Since most of the works in the literature focus on signal-degrading speckle, the information regarding signal-carrying speckle analysis is diffuse and sometimes abstruse. Thus, this review intends to provide a comprehensive overview of the different methods used to retrieve information from OCT speckle in biomedical applications. 

The remainder of this paper is structured as follows: Section \ref{sec:introduction} presents the introduction; Section \ref{sec:methods} presents the literature search criteria; Section \ref{sec:results} presents the signal processing methods used for analysing the OCT signal-carrying speckle. Discussion of some of the most relevant approaches is presented in Section \ref{sec:discussion}. Finally, the conclusions are presented in Section \ref{sec:conclusion}. Supplementary materials are provided in the Appendix, namely a table summarizing the reviewed works (Appendix \ref{sec:articlesdetails}) and a short theoretical mathematical description of light speckle (Appendix \ref{sec:speckletheory}).

\section{Methods}\label{sec:methods}

\begin{figure}[ht]
\centering
\includegraphics[width=0.7\textwidth]{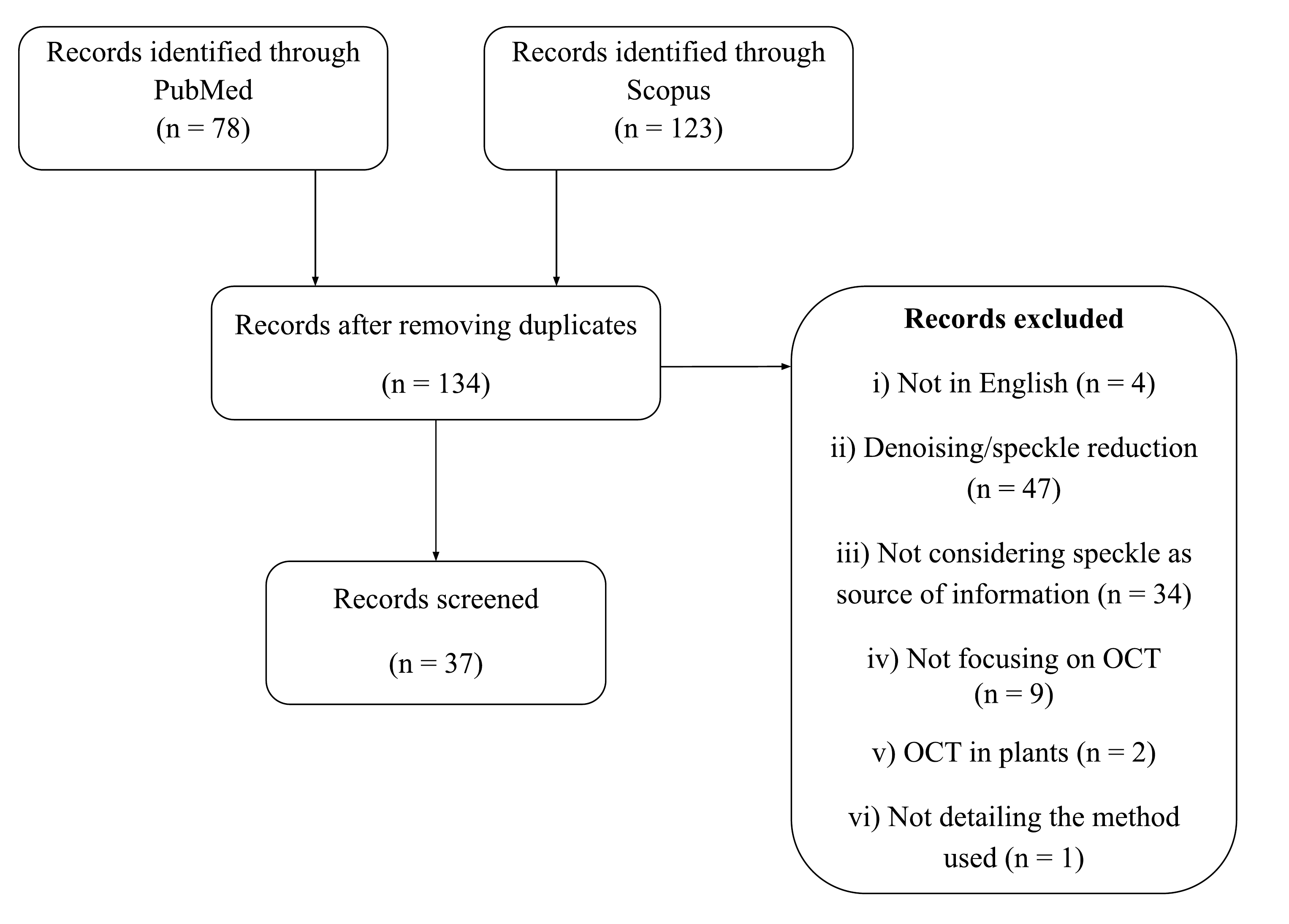}
\caption{Flowchart of the records selection.}
\label{fig:flowchart}
\end{figure}

The literature search was conducted in two databases on April~3\textsuperscript{rd}, 2021. PubMed was chosen for being one of the largest databases in the medical field, and Scopus for combining articles from both medical and technical fields. The search query used was: \emph{``Optical Coherence Tomography" AND speckle AND (statistics OR statistical) NOT flowgraphy}. After duplicate removal, the total number of articles obtained was 134. These articles were screened, and narrowed down to 37. The applied exclusion criteria were: i) not written in English, ii) focusing on denoising/speckle reduction, iii) not considering speckle as a source of information, iv) not focusing on OCT, v) OCT used in plants, vi) not detailing the method used. The number of articles excluded by each criterion is detailed in Figure \ref{fig:flowchart}. The remaining 37 articles were then reviewed.

The data extracted from each article were: the implemented method, the OCT technique used, the light-source wavelength (for non-theoretical studies), the biomedical application, and performance metrics related to the application, when provided. This information is reported in Table \ref{table:articles}, in Appendix \ref{sec:articlesdetails}.

\section{Results}\label{sec:results} 

\begin{figure}[ht]
\centering
\includegraphics[width=0.5\textwidth]{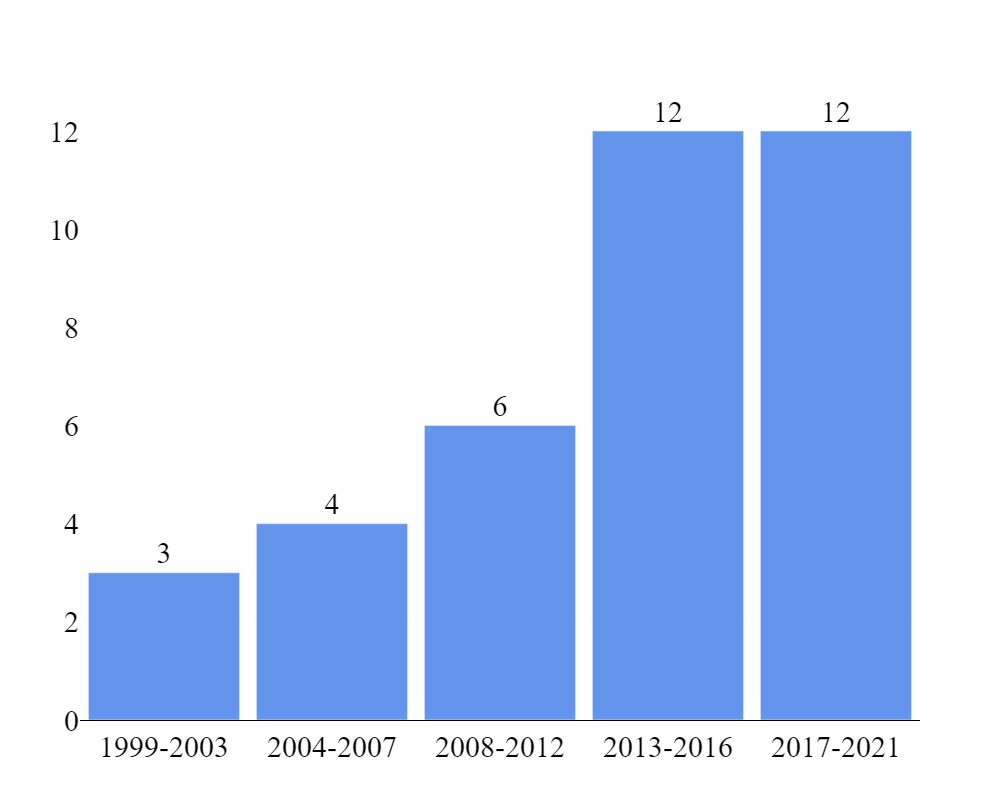}
\caption{Distribution of the articles included in the review, grouped by publication year.}
\label{fig:articles}
\end{figure}

Figure \ref{fig:articles} shows the distribution of the articles included in the review grouped by year. The results show a growing interest in the analysis of signal-carrying speckle in OCT imaging over the last two decades.

\begin{figure}[ht] 
\centering
\includegraphics[width=0.75\textwidth]{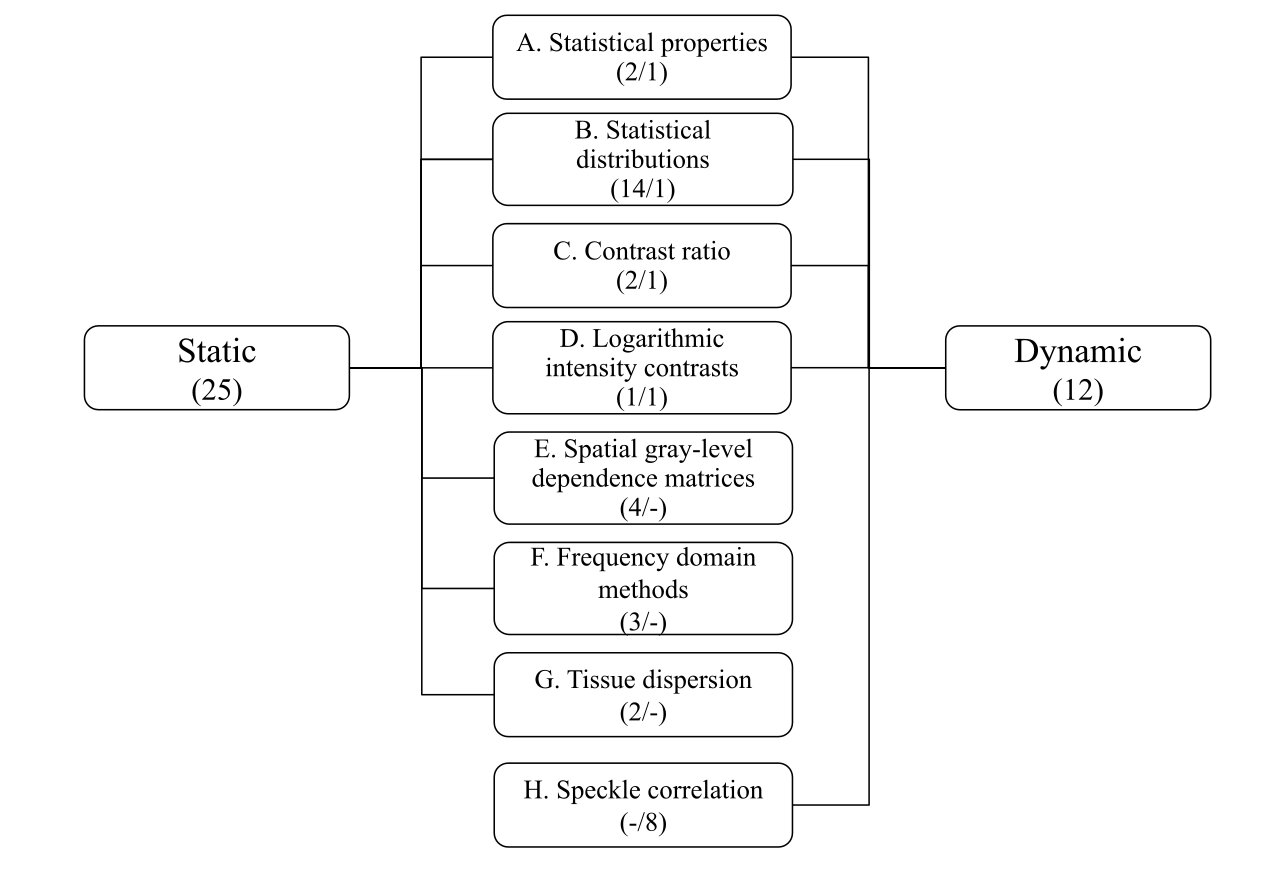}
\caption{Methods for speckle analysis included in this review. The number of articles included for each method performing a static ($s$) and dynamic ($d$) analysis is included as ($s$/$d$).}
\label{fig:methods}
\end{figure}

Speckle pattern analysis can either provide static or dynamic information about the imaged tissue, depending on whether the scatterers are stationary or in motion between consecutive image acquisitions. In this review, 12 articles were found performing a dynamic analysis, whereas the remaining 25 performed a static analysis. Figure~\ref{fig:methods} clusters the reviewed articles by technique, depicting the organization of Section~\ref{sec:results}.

\subsection{Statistical properties}\label{subsec:statprop}

Local moment based statistical properties of OCT signal intensity have been used for inferring speckle characteristics. These properties have proven to be useful in classification tasks, allowing to discriminate between different types of tissue. \\

\subsubsection{Static analysis}
Roy \textit{et al.} \cite{roy2015bag} used the mean ($\mu$), standard deviation ($\sigma$), kurtosis ($\kappa$), skewness ($\nu$), and an estimate of optical attenuation and signal confidence measures to detect tissue's susceptibility to rupture using intravascular OCT, with the objective of assessing atherosclerosis. The features proved to have high performance in the identification of such tissues using a Random Forest predictive model (area under the receiver operating characteristics curve of 0.9676).

Wang \textit{et al.} \cite{wang2013three} implemented a model for the detection of soft tissue sarcomas in  OCT images of \textit{ex vivo} human tissues, also based on speckle statistical properties. Specifically, the standard deviation of the signal fluctuations (speckles) of a single axial line (A-scan) was used. The statistical analysis of the features showed they were effective for comparing normal fat tissue and soft tissue sarcoma  (\textit{p-value}$<$0.01, Student's t-test). \\

\subsubsection{Dynamics Analysis of Speckle}
\begin{figure}[ht] 
\centering
\includegraphics[width=0.5\textwidth]{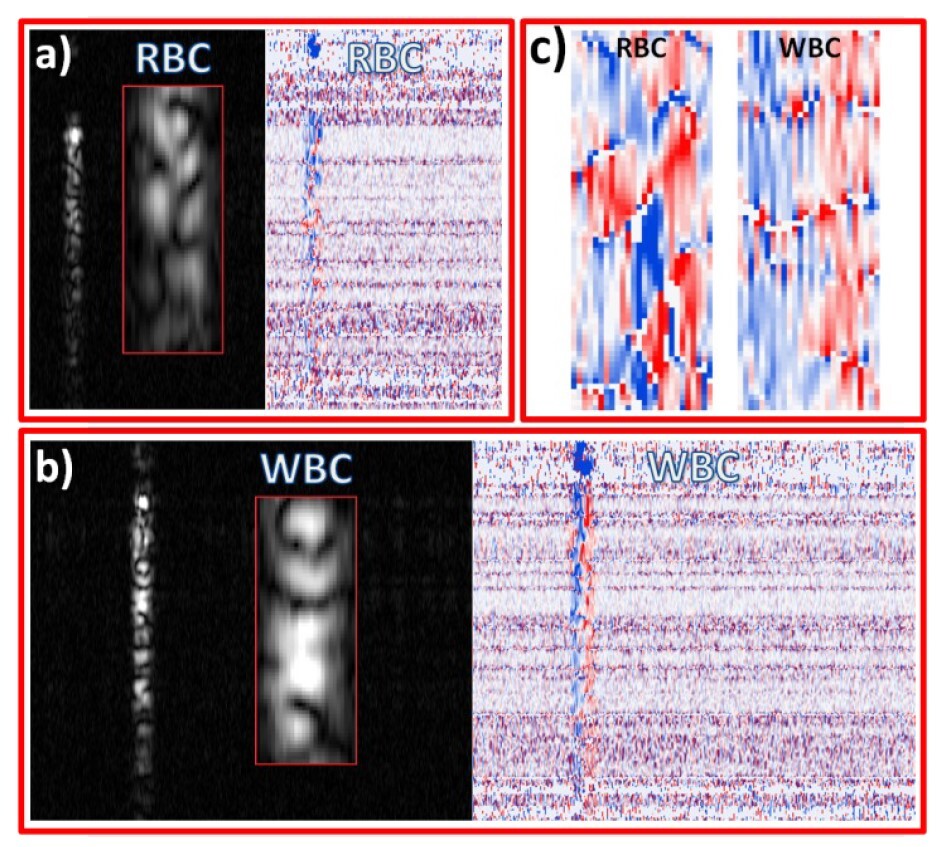}
\caption{Intensity and phase-change images originated from: a) modulation signal originated from erythrocytes (red blood cells, RBC) and b) modulation signal originated from leukocytes (white blood cells, WBC). In c) an enlarged subsection of RBC and WBC phase-change images, containing entire signals transversely, is shown. Reproduced from Ossowski \textit{et al.} \cite{ossowski2015detection} with the authors' permission.}
\label{fig:ossowski}
\end{figure}

Ossowski \textit{et al.} \cite{ossowski2015detection} used statistical properties of the OCT speckle to infer the dynamic properties of blood samples. Specifically, they used the mean horizontal and vertical speckle sizes, calculated from intensity data, and the sum of standard deviations of selected windows, calculated from phase data. These three statistical parameters were computed in OCT images of blood samples (Figure \ref{fig:ossowski}a and \ref{fig:ossowski}b), enabling a visual distinction between the signal modulation from erythrocytes and leukocytes as shown in Figure \ref{fig:ossowski}.

\subsection{Statistical distributions}

In Appendix \ref{sec:speckletheory}, the theoretical distributions for speckle complex amplitude (Gaussian distribution) and intensity (exponential distribution) are presented. These distributions have a clear physical meaning, but they are only applicable to an ideal, fully developed, speckle pattern. This may not always be the case in real world applications, where speckle formation could be difficult to model. Therefore, different probability density functions (PDFs) have been proposed to describe the speckle statistics in real world applications of OCT imaging. The parameters of these distributions are expected to change according to the light source properties and dimension/organization of the scatterers in the sample, thus providing information about the tissue properties. Given the different notations and formulations in the literature, a coherent mathematical notation of the proposed models is provided.

This section is organized as follows: in subsection \ref{subsubsec:fundamental}, Rayleigh and K-distribution are presented. These are the two fundamental distributions used to represent fully and non-fully developed speckle patterns, respectively. In subsection \ref{subsubsec:Gamma}, the Gamma and Generalized Gamma distributions are detailed. From these two, the remaining distributions, presented in subsection \ref{subsubsec:Gammaderived}, can be derived, including Weibull, Nakagami, Rician, three-parameter Rayleigh, and Lognormal. Finally, subsection \ref{subsubsec:dynamic} details the distributions that have been applied in a dynamic analysis.

\subsubsection{Fundamental distributions}
\label{subsubsec:fundamental}
The Rayleigh distribution is a 1-parameter distribution, used to model fully developed OCT speckle patterns.
The application of this model is valid when the signal arises from multiple scatterers within the resolution of the system \cite{mcheik2008speckle} and the light complex field amplitude is represented by circular Gaussian statistics, \textit{i.e.}, a fully developed speckle pattern \cite{lee2011speckle}. Its PDF is given by:

\begin{equation}
\label{eq:rayleigh}
    p_{RL}(A;a)=\frac{A}{a^{2}} e^{\left(-\frac{A^{2}}{2a^{2}}\right)} \: ,
\end{equation}

\noindent
where $a$ is the scale parameter.

Almasian \textit{et al.} \cite{almasian2017oct} experimentally verified the goodness of fit of the Rayleigh PDF for modelling speckle amplitude using controlled samples of silica microspheres suspended in water. They proved that OCT amplitude distribution for homogeneous samples can be described by a Rayleigh distribution for images with low optical depth (coefficient of determination, $R^{2}$ $\approx$ 0.98). Also, assuming a Rayleigh distribution, expressions were analytically derived for speckle signal mean, amplitude, and variance in terms of sample optical properties.

The K-distribution is a 3-parameter distribution, used to model cases where a small number of scatterers are present in the sample \cite{jesus2017assessment}, resulting in partially or non-fully developed speckle patterns. Its PDF can be written as:

\begin{equation}
\label{eq:kdist}
    p_{K}(A;\upsilon,\varphi,L)=\frac{2\xi^{(\beta+1)/2}A^{(\beta-1)/2}}{\Gamma(L)\Gamma(\varphi)}K_{\varphi-L}(2\sqrt{\xi A})  \; ,
\end{equation}

\noindent where $\beta=L+\varphi-1$, $\xi=L \varphi/\mu$, $K_{\alpha}$ is a modified Bessel function of the second kind of order $\alpha$. This distribution is the combination of two gamma distributions, one with mean 1 and shape parameter $\varphi$ and the other with mean $\upsilon$ and shape parameter L.

The K-distribution was tested by Jesus \textit{et al.} \cite{jesus2017assessment} for corneal OCT speckle intensity characterization against other distributions. However, using a Kolmogorov-Smirnov (KS) goodness of fit with 95\% confidence level, K-distribution modelled data presented statistical differences from original raw data, showing that it is not an adequate fit for the analysed problem. \\

\subsubsection{Gamma distributions}
\label{subsubsec:Gamma}

The Gamma distribution is a 2-parameter distribution. Its PDF belongs to a family of PDFs with two degrees of freedom, and is defined as \eqref{eq:gamma_pdf}:

\begin{equation}
\label{eq:gamma_pdf}
    p_{G}(A;a,d)=\frac{A^{d-1}e^{-A/a}}{a^{d} \Gamma(d)} \: \text{for} \: a,d>0 \; ,
\end{equation}

\noindent where $d$ is the shape parameter, $a$ is the scale parameter, and $\Gamma$ represents the Gamma function \cite{artin2015gamma}.

Kirillin \textit{et al.} \cite{kirillin2014speckle} developed a Monte Carlo model for speckle statistic simulation of OCT data, and validated the model using a phantom. Also, they demonstrated by visual inspection that the Gamma distribution was a good fit for both phantom and the previously simulated data. The scale parameter, $a$, showed an increase with the increase of scatterers concentration, whereas the shape parameter, $d$, presented a concentration-independent behavior. 

More recently, Niemczyk \textit{et al.} \cite{niemczyk2021effect} used the Gamma distribution to model speckle from corneal OCT data.
Both Gamma parameters showed a statistically significant relation with intra-ocular pressure (IOP) (\textit{p-value} $<$ 0.001, ANOVA test). 

The Generalized Gamma (GG) distribution is a 3-parameter generalization of the Gamma distribution, with a PDF given by:  

\begin{equation}
\label{eq:gg}
    p_{GG}(A;a, d, p) = \frac{p A^{d-1}}{a^d \Gamma(d / p)} e^{-(A/a)^{p}} \: \text{for} \: p>0 \;,
\end{equation}

\noindent where $d$ and $p$ are shape parameters, and $a$, the scale parameter. To obtain the Gamma PDF (Equation \eqref{eq:gamma_pdf}), $p$ must be set to 1. Special cases of GG include the previously presented Rayleigh (Equation \eqref{eq:rayleigh}), by setting the parameters to $p_{GG}(A;a\sqrt{2},2,2), a>0$.

\begin{figure}[h] 
\centering
\includegraphics[width=0.6\textwidth]{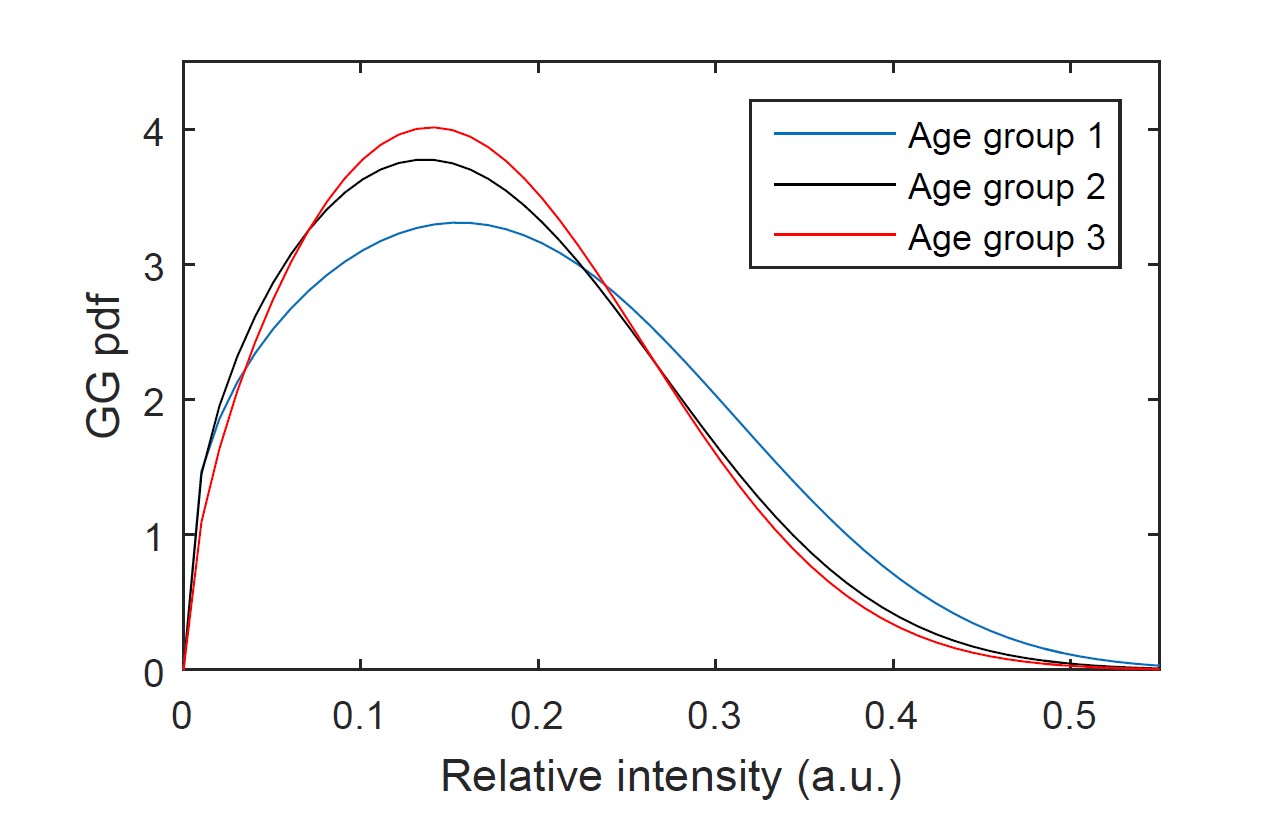}
\caption{Probability density function of the Generalized Gamma (GG) distribution for three different age groups, where Age group 1 is the youngest and Age group 3, the oldest. Reproduced from Jesus \textit{et al.} \cite{jesus2017assessment} with the authors' permission.}
\label{fig:GG_ages}
\end{figure}

The GG distribution was used by Jesus \textit{et al.} \cite{jesus2017influence,jesus2015age,jesus2017assessment} and Iskander \textit{et al.} \cite{iskander2020assessing} to model corneal OCT data. Jesus \textit{et al.} \cite{jesus2015age,jesus2017assessment} applied the GG distribution to healthy subjects divided in three age groups (24.4$\pm$0.5, 31.3$\pm$4.6; 61.2$\pm$8.4 years), as depicted in Figure \ref{fig:GG_ages}. The goal was to study variations of the distribution parameters among groups. A significant statistical difference, (\textit{p-value}$<$0.05, Kruskal-Wallis test) was observed for all three parameters.

In a later study \cite{jesus2017influence}, Jesus \textit{et al.} analysed the parameters' relation with micro-structural corneal properties. Significant correlation (\textit{p-value}$<$0.001) was found between both the scale parameter ($a$) and the ratio of the shape parameters ($d/p$) with intraocular pressure (IOP). The authors suggested that GG parameters can contribute to improve IOP measurements.

Iskander \textit{et al.} \cite{iskander2020assessing}, used the GG to model information from the micro-structure of the cornea to differentiate glaucoma suspects, glaucoma patients, and healthy controls. ANOVA tests showed that the scale parameter, $a$, was correlated with the shape parameter, $p$, and the relation between these two parameters was statistically significantly different between the three study groups (\textit{p-value}$<$0.0001, Fisher's Test). 

Finally, Seevaratnam \textit{et al.} \cite{seevaratnam2014quantifying} used the GG distribution to investigate the effect of temperature variation in tissue phantoms. The scale parameter, $a$, showed a linear increase with the increase of the tissue temperature. The correlation between $a$ and the temperature was statistically significant (\textit{p-value}=$7.9\times10^{-6}$, Student's t-test).\\

\subsubsection{Gamma derived distributions}
\label{subsubsec:Gammaderived}

The 2-parameter Weibull distribution \cite{jesus2017assessment}, can be obtained from GG distribution for $d=p$:

\begin{equation}
    p_W(A;a,d) = \frac{d A^{d-1}}{a^d} e^{-(A/a)^d} \: .
\end{equation}

Jesus \textit{et al.} tested this distribution to model speckle corneal intensities \cite{jesus2017assessment}. No statistically significant difference was observed between the fitted and the raw data (\textit{p-value} $<$ 0.05, KS goodness of fit test), concluding that the Weibull distribution can be used to model these data.

The Nakagami distribution is a 2-parameter distribution that can be obtained from the Gamma distribution by setting $a=\Omega/d$ and taking the square root of the original random variable, $A'=\sqrt{A}$ \cite{huang2016nakagami}.

\begin{equation}
\label{eq:nakagami}
    p_{NK}(A';d,\Omega)=\frac{2d^d}{\Gamma(d)\Omega^d}A'^{2d-1}e^{-\frac{d}{\Omega}A'^2} \; ,
\end{equation}

\noindent where $d$ is a shape parameter and $\Omega$ is a spread parameter.

This distribution has been proposed to represent the dispersion of several backscattered clusters of incoherently added waves \cite{mcheik2008speckle}, and has been tested for modeling skin speckle data against other distributions. It was considered the best fit, using the Kolmogorov-Smirnov test, and its parameters showed statistical differences between two different skin layers (Epidermis and Stratum Corneum). 

Nakagami distribution was also tested in corneal data \cite{jesus2017assessment,jesus2015age}. For this case, although it was not considered the best fit, fitted data did not present statistical significant differences from the raw data, also using the KS test.

The Rician, or Rice distribution (Equation \eqref{eq:Rician}) is a 2-parameter generalization of the Rayleigh distribution (Equation \eqref{eq:rayleigh}), obtained by the introduction of a noncentrality parameter:

\begin{equation}
\label{eq:Rician}
    p_{RI}(A;a,\nu)=\frac{A}{a^{2}} e^{-\frac{A^{2}+\nu^{2}}{2a^{2}}} I_0 \left( \frac{A\nu}{a^{2}}\right) \; ,
\end{equation}

\noindent where $\nu$ is the noncentrality parameter and $I_{0}$ is the zero order modified Bessel function of the first kind \cite{weisstein2002bessel}. Thus, the Rayleigh can be obtained from the Rician distribution for $\nu = 0$.

The validity of this distribution for modelling speckle amplitude distribution was tested for tissue phantom data by Seevaratnam \textit{et al.} \cite{seevaratnam2014quantifying} and for corneal data, by Jesus \textit{et al.} \cite{jesus2017assessment}. While it was not considered the best fit for the tissue phantoms, it was able to model the data successfully. However, for the corneal data, the data modelled with this distribution presented statistically significant difference from raw data (Kolmogorov-Smirnov test for a 95\% confidence level).

The 3-parameter Rayleigh distribution is obtained by modifying the Rayleigh (Equation \eqref{eq:rayleigh}) including two new parameters, $b$ and $c$:

\begin{equation}
\label{eq:rayleigh3}
    p_{3RL}(A;a,b,c)=\frac{b(A-c)}{a^{2}}e^{-\frac{-(A-c)^{2}}{2a^{2}}} \; ,
\end{equation}

\noindent with $a$ the scale parameter, $b$ the amplitude normalization parameter, and $c$ the shifting parameter.

Matveev \textit{et al.} \cite{matveev2019oct} and Demidov \textit{et al.} \cite{demidov2019analysis} used spatial speckle statistics on OCT lymphangiography and neurography to map lymphatic vessels, based on the analysis of the parameters of $p_{3RL}$. Their experiments, on normal skin and tumor tissues, showed that, by fitting Equation \eqref{eq:rayleigh3} to different regions of interest (ROIs) in an image, the obtained $R^{2}$ values statistically differed from each other, and could then be used as a feature for nerves and lymphatic vessels mapping. By using a threshold on the $R^{2}$ value ($0.9<R^2<0.99$ for the lymph vessels), the authors were able to obtain a discrimination of the tumor from the normal tissue. Following their previous studies, Matveev \textit{et al.} \cite{matveev2019assessment} presented an optimization model to automatically determine the threshold for the $R^{2}$ value and for the size of the ROI, both parameters previously empirically chosen.

The Lognormal distribution is a 2-parameter distribution of a variable whose logarithm follows a normal distribution, with mean $\upsilon$ and standard deviation $\sigma$. Its PDF is given by Equation \eqref{eq:lognormal}, and can be derived from the GG distribution (Equation \eqref{eq:gg}) by setting $d/p \rightarrow \infty$:

\begin{equation}
\label{eq:lognormal}
    p_{L}(A;\mu,\sigma)=\frac{1}{\sigma A \sqrt{2\pi}} e^{-\frac{(log A-\upsilon)^{2}}{2\sigma^{2}}} \; ,
\end{equation}

The Lognormal distribution has been applied by Jesus \textit{et al.} \cite{jesus2017assessment} and Mcheik \textit{et al.} \cite{mcheik2008speckle}, on corneal and skin speckle data, respectively. Both studies have been further described in Subsection \ref{subsubsec:Gamma}, as the authors compare the Lognormal distribution with the GG distribution in both cases. They also obtain the same conclusion: the GG distribution is a better fit than the Lognormal for corneal and skin speckle data, and data modelled with Lognormal distribution show statistically significant difference from the original data, using the Kolmogorov-Smirnov test with a level of significance of 0.05.\\

\subsubsection{Statistical Distribution in Dynamics Analysis of Speckle}\label{subsubsec:dynamic}

The temporal speckle distribution of a single pixel is expected to follow different distributions according to the properties of the sample in that pixel. Therefore, a statistical distribution can be applied to the time-domain histogram of individual pixels to account for speckle dynamics.  

Cheng \textit{et al.} \cite{cheng2014histogram} analysed OCT voxels denoting fluid flow for large and small arterioles and venules in phantom and skin data. This analysis was performed as part of a visualization enhancement technique. The authors state that a pixel located within static tissue is expected to follow a Gaussian distribution over time, while pixels located in regions depicting flow will follow a different distribution. In their results, they concluded that Rayleigh distribution (Equation \eqref{eq:rayleigh}) is suitable to describe the speckle temporal distribution of large arterioles and venules, while the Rician distribution (Equation \eqref{eq:Rician}) can model the amplitude of OCT for tissues that are partially denoting static and flowing scatterers, such as capillaries.

\subsection{Contrast ratio}
\label{subsec:contrast}
The contrast ratio (C) of an OCT image can be defined as the ratio of the signal’s standard deviation, $(\sigma)$, and its mean, $(\mu)$ \cite{hillman2006correlation},

\begin{equation}
\label{eq:contrast}
    C=\frac{\sigma}{\mu}
\end{equation}

\noindent The contrast ratio is expected to vary with the density of scatterers, and has been used for different applications, such as segmentation and motion estimation.

\subsubsection{Static analysis}

Hillman \textit{et al.} \cite{hillman2006correlation} and Duncan \textit{et al.} \cite{duncan2008statistics} both proved that it is possible to obtain a correlation between the local contrast statistics and the scatterers density in an OCT sample. Hillman \textit{et al.} \cite{hillman2006correlation} theoretically demonstrated that contrast ratio decreases with the increase of the effective number of scatterers that contribute to the signal. The authors also confirmed empirically their predictions by using an experimental set-up of controlled tissue phantoms of suspensions of microspheres in water with different concentrations. Duncan \textit{et al.} \cite{duncan2008statistics} computed the local contrast image of simulated synthetic speckle patterns. Then, they estimated the relation between the lognormal distribution (Equation \eqref{eq:lognormal}) parameters of this image, and the size of the image kernel. Experiments using chick embryo OCT images were also performed. Vessels and background were successfully segmented by choosing an adequate threshold of the lognormal PDF parameters computed in the contrast image.\\

\subsubsection{Dynamics analysis of speckle}

Kirkpatrick \textit{et al.} \cite{kirkpatrick2007quantitative} developed an approach to quantify the shift and temporal contrast in a translating speckle pattern. The end goal of the authors was to quantify local motion. Their proposed method, quantitative temporal speckle contrast imaging, is dependent on the speckle size in the image and the number of images in a sequence. The application of the method is also depends on the acquisition speed of the device, which should be fast enough that no motion occurs during the image acquisition.

\subsection{Logarithmic pixel intensity contrasts}\label{subsec:logintcon}

OCT images are often displayed in a logarithmic scale to enhance dynamic ranges. This transformation causes some properties to vary from the OCT data that has not been logarithmically transformed.\\

\subsubsection{Static analysis}
Considering a single image where amplitudes follow the Rayleigh distribution (Equation \eqref{eq:rayleigh}), its logarithmic transformation will result in a contrast ratio defined in Equation \eqref{eq:contrast}, a function inversely dependent on the logarithmically transformed mean intensity signal of the signal intensity \cite{lee2011speckle}. In these conditions, the speckle contrast is equivalent to:

\begin{equation}
\label{eq:contrastlog}
    C=\frac{\pi}{\sqrt{6} (D_{I} - \gamma)} \; ,
\end{equation}

\noindent where $\gamma$ is the Euler-Mascheroni constant, and $D_{I}$ is the logarithmically transformed mean intensity signal. In an OCT acquisition, the light beam propagates through the tissue from top to bottom, which causes its intensity to decrease with the increase of the depth. Then, it is expected that speckle contrast of the logarithmically transformed OCT signal will also be dependent on depth. However, the level of signal attenuation will also depend on the scattering properties of the tissue. 

Lee \textit{et al.} \cite{lee2011speckle} used simulated OCT images and \textit{in vitro} rat liver images to explore these properties, and demonstrate the validity of the relation in Equation \eqref{eq:contrastlog}. Moreover, they showed that the contrast value changes according to the scattering coefficient of the sample, which can be used to characterize tissues.\\

\subsubsection{Dynamics analysis of speckle}

The logarithmic intensity variance (LOGIV) is defined as the variance of the intensity image after the logarithmic transformation. The differential logarithmic intensity variance (DLOGIV), is calculated by multiplying the LOGIV by a factor of two. LOGIV and DLOGIV values approach $\pi^2/6$ and $\pi^2/3$, respectively, when the signal-to-noise ratio (SNR) approaches zero. These values can be used to compute LOGIV and DLOGIV tomograms by collecting multiple scans from the same region and measuring the quantitative variance of logarithmic intensities over scans. 

Motaghiannezam \textit{et al.} \cite{motaghiannezam2012logarithmic} proposed this technique for the analysis of \textit{in vivo} human retinal vasculature visualization. Results of their experiments showed that static areas of the retina were invisible in the LOGIV and DLOGIV tomograms, while areas with detectable motion, such as blood vessels, were not. They also confirmed the low sensitivity of LOGIV and DLOGIV to the sample reflective strength, demonstrating the superiority of these methods in comparison to linear contrast ratios for detecting motion and visualizing microvasculature.

\subsection{Spatial Gray-Level Dependence Matrices (SGLDM)}
\label{subsec:SGLDM}

\begin{figure}[t]
\centering
\includegraphics[width=0.7\textwidth]{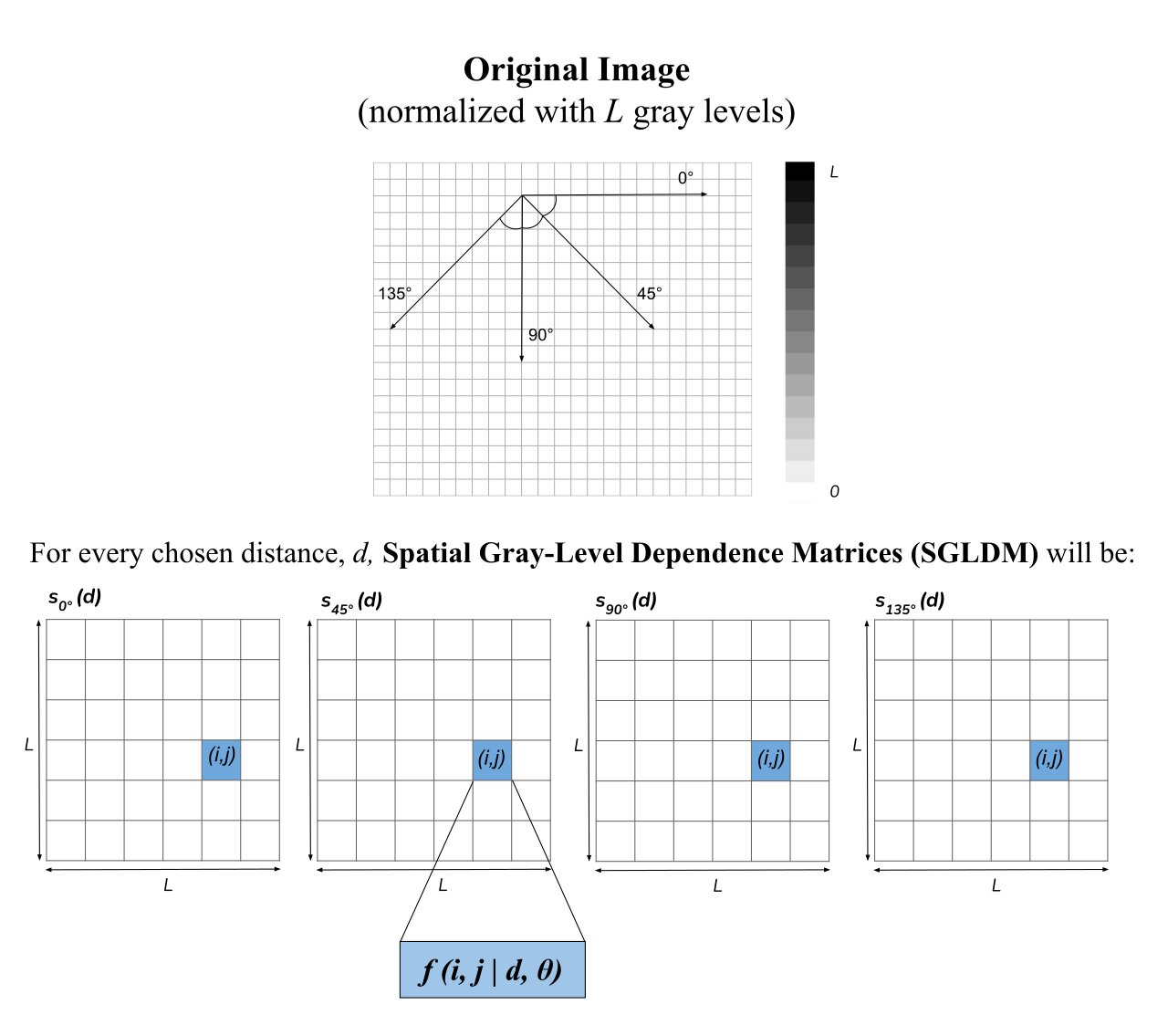}
\caption{Diagram representing the final result of SGLDM where \textit{$s_{\theta}$ (i,j $\mid$ d)} is the SGLDM matrix for distance \textit{d} and direction $\theta$, and \textit{f(i,j $\mid$ d,$\theta$)} represents the probability of a pixel with level \textit{i} being at a distance \textit{d} from a pixel with gray level of \textit{j} in the $\theta$ direction.}
\label{fig:sgldm}
\end{figure}

Spatial gray-level dependence matrices (SGLDM), also referred to as co-ocurrence matrices \cite{gossage2003texture, gossage2003texture2,gossage2006texture,kasaragod2010speckle}, are determined by the estimation of the second-order joint-probability distribution of each combination of grey-level values that occur next to each other, averaged over directions of 0º, 45º, 90º, and 135º. Assuming the images are normalized with \textit{L} grey scale levels, each \textit{f(i,j $\mid$ d,$\theta$)} is the probability of a pixel with level \textit{i} being at a distance \textit{d} from a pixel with gray level of \textit{j} in the $\theta$ direction. Following this principle, an $L \times L$ matrix can be created for each direction and for each chosen distance, as is shown in Figure \ref{fig:sgldm}. 

Several features can be extracted from these matrices, including energy, entropy, correlation, local homogeneity, and inertia, detailed in \cite{gossage2003texture,gossage2003texture2,gossage2006texture}. Similarly to the statistical properties described in subsection \ref{subsec:statprop}, these features are expected to be linked to changes in tissue. Thus, they can be used for classification tasks.

Gossage \textit{et al.} \cite{gossage2003texture} proposed and applied SGLDM features 
for differentiating mouse skin and fat, and normal versus abnormal mouse lung tissue, using a minimum-error-rate Bayesian model. Their results showed a high accuracy classification of mouse skin and fat, of 98.5 and 97.3\%, respectively. A satisfactory performance was also obtained for distinguishing normal and abnormal mouse lung, of 64.0 and 88.6\%, respectively. Finally, the features were used to classify five different bovine tissues, with a similar classification model, resulting in an average of 80\% correct classification rate. 

Kasaragod \textit{et al.} \cite{kasaragod2010speckle} used SGLDM to retrieve information from the speckle OCT images. A Bayesian model was applied to the classification of tissue phantoms with different amount of scatterers, and to identify the invasion of melanoma cell into tissue engineered skin. Their results were satisfactory in classifying the number of scatterers in the tissue phantoms, shown visually by a ROC curve plot. However, this approach provided limited results in the identification of the melanoma cells in the tissue. 

\subsection{Frequency Domain methods}
\label{subsec:frequency}

After computing the 2D Discrete Fourier Transform (DFT) in an OCT image, the resulting image can be divided into regions, according to their frequency content. This will result in different texture parameters per region. The contribution of each region to the total frequency magnitude is calculated by summing all the values of the spatial frequencies in that region, and dividing by the total frequency magnitude of the image. This value represents the percentage of signal within a certain range of spatial frequencies, and it can be used as a feature in similar applications as described in \ref{subsec:statprop} and \ref{subsec:logintcon}.

Gossage \textit{et al.} \cite{gossage2003texture, gossage2003texture2} used the 2D-DFT for retrieving information from OCT images, together with the previously described SGLDM (subsection \ref{subsec:SGLDM}). The goal of the work was analysing and classifying texture of different tissues (mouse lung and bovine tissue). The results detailed in section \ref{subsec:SGLDM} were obtained with a combination of 2D-DFT and SGLDM features. A similar approach was used in \cite{gossage2006texture}, where the 2D-DFT and SGLDM features were used to differentiate living from non-living tissue phantoms with various sizes and distributions of scatterers. 

\subsection{Tissue dispersion}

Tissue dispersion can be measured in a static image analysis from the degradation of the image Point Spread Function (PSF). The standard method to measure tissue dispersion is through the computation of the resolution degradation of single reflections. However, such a measurement requires distinct point reflectors, below and outside the sample, which rarely happens \textit{in vivo}. 

As alternative, Photiou \textit{et al.} \cite{photiou2017measuring,photiou2017using} proposed to estimate the tissue dispersion from the imaged speckle. Being a coherent phenomenon, speckle is affected by tissue dispersion. Changes in speckle size can be used to estimate the broadening of the image PSF, and later to calculate the group velocity dispersion (GVD). This method is based on the comparison of small regions of an OCT image at different depths without visible structures. These regions should only contain speckle information and it is expected that a region from the surface shows no dispersion, as opposed to deeper regions.
Photiou \textit{et al.} \cite{photiou2017measuring,photiou2017using} showed that their proposed method performs similarly to the standard procedure (from degradation of the image PSF), with a GVD difference less than 7\%. Also, this value proved to be a good feature for tissue classification, when comparing normal to malignant samples of human colon (accuracy of 96\% using linear discriminant analysis).

\subsection{Speckle correlation}

Considering a sequence of scans over time, the autocorrelation and/or decorrelation of intensities can carry information about motion of particles in a sample. Both features can be used as an approach for time-varying speckle analysis \cite{de2015new}. 

In terms of speckle analysis, the normalized autocorrelation, for a given point $p$ and time lag $\tau$, is given by:

\begin{equation}
\label{eq:corr}
   g_{p,\tau}=\frac{\sum_{t=0}^{N-1} (I_{p,t}-\langle I_p \rangle) (I_{p,t+\tau}-\langle I_p \rangle)}{\sum_{t=0}^{N-1} (I_{p,t}-\langle I_p \rangle)^2} \; ,
\end{equation}

\noindent where $I_{p,t}$ is the intensity in the point $p$ at time $t$, $\tau$ is the period of time between scans, and $N$ is the total number of scans considered. The temporal average intensity, $\langle I_p \rangle$, is subtracted from each intensity value in order to consider only the intensity fluctuations.
The autocorrelation of a speckle pattern with itself ($\tau=0$) is expected to be maximized. With the increase of the lag between scans, this value is expected to decrease until it reaches 0, when the scans are no longer correlated. A vector of autocorrelation values can be obtained by changing $\tau$. These autocorrelation values are related with the fluctuations in the intensity of the speckle patterns. At the same time, the fluctuations are related with the flow of particles in the tissue, which in turn are expected to be related with the flow of particles in the tissue \cite{de2015new}. 

The decorrelation time $\tau_c$ is used to analyse the shape of each autocorrelation function \cite{de2015new}. This metric corresponds to the time it takes for the autocorrelation value to fall to $1/e$. 

  
De Pretto \textit{et al.} \cite{de2015new} performed experiments with milk pumped through a microchannel at different velocities, and proved the inversely proportional relation between decorrelation time and flow velocity. As expected, this relation is highly dependent on the sampling frequency. A sampling rate of 8kHz, makes the system appropriate for differentiating between low flow rates, up to $12 \; \mu l/min$. However, if the flow rates are higher, the temporal resolution of the system makes it unsuitable.

In a later study, De Pretto \textit{et al.} \cite{de2016optical} implemented the same approach to monitor blood sugar in OCT data. They used samples of heparinized mouse blood, phosphate buffer saline, and different concentration of glucose. 
They were able to differentiate between low level of glucose concentration, up to 355 mg/dL, indicating the suitability of OCT for non-invasive measurements of glucose levels. 

Popov \textit{et al.} \cite{popov2017statistical} conducted an experiment using tissue phantoms. They obtained an expression for the spatio-temporal correlation function of scattered radiation, assuming a single scatterer regime, and were able to accurately measure viscosity.


Farhat \textit{et al.} \cite{farhat2011detecting} assessed the changes that occur in intracellular motion as cells undergo apoptosis. To that end, they induced apoptosis in samples of acute myeloid leukemia cells, and they measured the decorrelation time of the speckle over a period of 48h. Their results showed an increase of motion in the cells (identified as a decrease of the decorrelation time of speckle) after 24h, which is in accordance with histology. 

Ferris \textit{et al.} \cite{ferris2020forward} used phantom data to study the effects of multiple scattering on the speckle decorrelation. Their conclusions confirm that speckle decorrelation is dependent on parameters such as the concentration and size of particles, and velocity field inhomogeneities. They also concluded that an overestimation of blood flow velocities might occur because an increase in the rate of decorrelation is caused by the detection of forward scattered light. 

Uribe-Patarroyo \textit{et al.} \cite{uribe2014quantitative} proposed a new discrete normalized second-order autocorrelation.

The authors claim this approach proves to be more robust to the presence of noise, and can be used as a method for speed measurements in tissue phantoms. This same method was used in a later work \cite{uribe2015rotational} for the correction of the rotation distortion in catheter-based endoscopic OCT.

Finally, Liu \textit{et al.} \cite{liu2013robust} used the cross-correlation coefficient between adjacent A-scans to analyse properties of simulated speckle images. Their results underline the importance of over-sampling when calculating motion properties from temporally dynamic speckle, such as the contrast or decorrelation time, due to its random nature.

\section{Discussion}\label{sec:discussion}

The reviewed approaches can be categorized as static or dynamic, depending on the type of data that is used, with some methods being applicable to both categories. In a static analysis, the inherent speckle pattern in a single image is studied, in order to obtain information about the micro-structures in the sample. In a dynamic analysis, the changes in speckle between images are analyzed to gain information on the dynamic nature of the moving scatterers \cite{vaz2016laser, vaz2017effect, sugiyama2010use}.

In most of the reviewed works, the authors aimed to further understand the physical meaning of the light speckle, and how to model it mathematically \cite{karamata2005speckle,liu2013robust,ferris2020forward,kirillin2014speckle, schmitt1999speckle,almasian2017oct,kirkpatrick2007quantitative,duncan2008statistics,lee2011speckle}. Although its interpretability remains a challenging task, a number of works have shown the feasibility of speckle-derived quantifications for biomedical imaging related tasks, including  classification (e.g. healthy/pathological) \cite{wang2013three,ossowski2015detection,roy2015bag, seevaratnam2014quantifying,jesus2015age,jesus2017assessment,jesus2017influence,demidov2019analysis,matveev2019assessment,matveev2019oct,iskander2020assessing,cheng2014histogram,niemczyk2021effect,photiou2017measuring,photiou2017using,kasaragod2010speckle,gossage2003texture,gossage2003texture2,gossage2006texture},  segmentation (e.g. vessels) \cite{mcheik2008speckle}, or motion quantification (when dynamic data is provided) \cite{de2015new,farhat2011detecting, liu2013robust, uribe2014quantitative,uribe2015rotational,ferris2020forward}. Nevertheless, some aspects and limitations of those techniques should be discussed in order to improve future works.

Among all the reviewed methodologies, speckle modeling using statistical distributions has been the most studied, especially on static data. Several authors have proven the applicability of these approaches to different tasks, mainly to classify between different tissues. However, the conclusions on the optimal PDF often vary depending on the analysed tissue, the application, and the OCT device. For example, Mcheik \textit{et al.} \cite{mcheik2008speckle} and De Jesus \emph{et al.} \cite{jesus2017assessment} both presented a comparative study including several PDFs, but in different applications, one to differentiate speckle from different skin layers, and the second from different groups in corneal data. Despite both works including Nakagami, GG, and Lognormal distributions, their conclusions were different. For skin layers, the Nakagami distribution was found to be the best. However, for corneal data, the best fit was achieved by the GG distribution. Furthermore, other authors have successfully applied other distributions to the same application, such as Niemczyk \textit{et al.} \cite{niemczyk2021effect}, who used the Gamma distribution to study the effect of IOP on corneal OCT speckle from porcine eyes. This precludes drawing conclusions on the PDF which provides the best fit for each application, since in most of them, exhaustive comparisons do not exist.

A drawback that the modeling of speckle through statistical distributions had to tackle is that the real world problems do not fulfill the theoretical assumptions of speckle formation. Theoretically speaking, when the number of elementary phasors is high, meaning a high number of scatterers per coherence volume, the central limit theorem is fulfilled \cite{lee2011speckle}, and the speckle pattern is fully developed, and its intensity distribution follows the Rayleigh distribution. However, this argument is only partially applicable to biological tissues, because different tissues may have different natures, \textit{i.e.} some can be more heterogeneous with a lower number of scatterers \cite{demidov2019analysis}. As a consequence, several authors propose more complex distributions (three parameters and higher), which presents new potential challenges. When a statistical model is used to represent the process that generated the data, the representation will not be fully accurate, as some information will be lost. In estimating the amount of information lost by a model, one needs to take into account the trade-off between the goodness of fit and the simplicity of the model, i.e. the risk of overfitting and underfitting. In the corneal studies \cite{jesus2015age,jesus2017assessment, iskander2020assessing}, the Akaike’s Information Criterion is applied to minimize this risk. 

Following the analysis of statistical distributions, the second most used technique is the analysis of the correlation, which is only applicable to dynamic data. The main application of this technique is in motion determination (e.g. blood flow). Although most of the reviewed works in OCT speckle are focused on theoretical modeling or validation with phantoms, this is in fact a type of signal analysis that has been widely explored in other imaging modalities. Some examples are laser speckle imaging, which can be used for cutaneous blood flow determination \cite{vaz2016laser, vaz2017effect}, blood pulse pressure waveform estimation \cite{vaz2015laser,beiderman2010remote},  cellular assessment in muscle tissue \cite{maksymenko2015application}, and laser speckle flowgraphy which can be used for ocular blood flow determination \cite{sugiyama2010use}. Furthermore, the rationale behind these techniques has also been widely applied to compute OCT angiography (OCTA) from a set of temporal OCT data acquired at the exact same position \cite{de2015review}.

In contrast with the correlation analysis, there have been little to none applications of the other methods for dynamic data. However, some authors have demonstrated the feasibility of statistical properties \cite{ossowski2015detection}, statistical distributions \cite{cheng2014histogram}, or logarithmic intensity contrast \cite{motaghiannezam2012logarithmic} to visualize and segment blood vessels within a tissue.

Finally, the methods that compute characteristics of static data (tissue dispersion, SGLDM, frequency domain methods, or the previously mentioned statistical properties/distributions and contrast (both with and without the application of the logarithmic transform)), are more applied to obtain features to use in a predictive model or classifier. While these approaches may under-perform classification models that take into account the complete image information or several features, many of these proposed features are very easy to interpret, and to link to the physical changes in the tissue, making them interesting for clinical practice.

Although a growing interest in the analysis of signal-carrying speckle has been observed over the last years, it is still a research line in an early stage. A considerable number of different methods have already been proposed but only a few applications published in the literature. At the current stage, the analysis of OCT speckle is lacking on validation and information on its reproducibility in-vivo. None of the reviewed studies validated the proposed methods on large (the largest dataset in the studies included was 65 \cite{jesus2015age,jesus2017assessment}) or multiple datasets. The speckle information is intrinsically related to the spatial arrangement and biomechanical properties of the scatterers in the sample. Scatterers can either be collagen fibers and fibroblasts when imaging the cornea, blood cells flowing through vessels, or just silica particles in a phantom image. What is considered a scatterer in a sample will depend on the characteristics of the imaging system, namely the relationship between the particle size and the light source wavelength \cite{siebert2000relationship}. 
This is particularly important for OCT imaging, given the variability existing between devices (790--1330nm), as it can be observed in Table \ref{table:articles}, Appendix \ref{sec:articlesdetails}. Consequently, for the same sample and method, different quantitative values for speckle may be obtained depending on the OCT system used. 

Another important aspect that hampers the development of speckle-based techniques is the limited access to raw data. Images collected from commercial OCT devices are often filtered to reduce the speckle or transformed to increase visualization contrast. It is of utmost importance that raw OCT images are used in speckle studies, otherwise the obtained results are tainted by the used device and pre-processing algorithm \cite{hu2013comparison}. If raw data is not available, the information of the applied image processing algorithms should be provided to understand how the speckle has been processed and hence, comprehend its physical meaning in a biomedical application. 

Despite its early stage, research on methods to study the signal-carrying speckle has been a step forward on the comprehension of the physical meaning of the information retrieved from OCT imaging. Speckle analysis provides information on the size and distribution of the scatters that has not been considered in a clinical practice yet. Such advancements are also particularly interesting for other research lines such as OCT elastography, adaptive optics imaging, or machine learning applications. For example, recent developments on machine learning, namely on convolutional neural networks  have reported outperforming results in OCT image analysis in comparison to conventional image processing \cite{fang2017automatic, raja2020extraction,tennakoon2017retinal}. However, deep learning approaches still lack on interpretability and roughly remain a black box, despite the recent efforts to address this limitation \cite{huff2021interpretation}. Therefore, future research may focus on integrating physics and learning based approaches, to combine their strengths.

\section{Conclusion}\label{sec:conclusion}

This paper presents an overview of the current state of the art in OCT signal-carrying speckle analysis in biomedical applications. The results of this literature review show that several methods have already been proposed for different applications, highlighting the potential of speckle analysis to infer the optical and spatial properties of the scatterers in a sample or tissue. However, signal-carrying speckle analysis in OCT is still in its early stage and further work is needed to validate its applicability and reproducibility in a clinical context.

\section*{Acknowledgments}

The authors thank Prof. Dr. D. Robert Iskander and Dr. Monika E. Danielewska for their technical revision of the paper, and respective contributions. The authors thank also Prof. Dr. Ingeborg Stalmans, Dr. João Breda and Dr. Jan Van Eijgen for their clinical advisory and support.

\section*{Disclosures}
The authors have no relevant financial interests in this article and no potential conflicts of interest to disclose.

\section*{Funding}
This work was supported by the Horizon 2020 Research and Innovation Programme (grant agreement no. 780989: Multi-modal, multi-scale retinal imaging project) and by Portuguese National Funds through the FCT, Fundação Para a Ciência e a Tecnologia, I.P., in the scope of the project UIDB/04559/2020.

\bibliographystyle{unsrt}
\bibliography{bibliography}

\clearpage
\newpage
\begin{appendices}

\newpage
\onecolumn
\global\pdfpageattr\expandafter{\the\pdfpageattr/Rotate 90} 
\begin{landscape}
\section{Articles Details}
\label{sec:articlesdetails}
\begin{singlespace}
\newcolumntype{C}[1]{>{\centering\arraybackslash}m{#1}}

\begin{longtable}{C{3.2cm} C{1.7cm}C{3cm}C{2.8cm}C{3.7cm}C{3.7cm}C{2cm}}

\caption{Characteristics of the reviewed studies}\\ \hline
\textbf{Authors \& Publication year} & \textbf{Static/ Dynamic} & \textbf{Method} & \textbf{Aim} & \textbf{Application/ Data used} & \textbf{OCT technique (brand)}  & \textbf{Light wavelength (nm)} \\ \hline
\endhead
Wang et al. \cite{wang2013three} (2013)                           & Static          & Statistical properties          & Classification          & Ex vivo human tissue                                                    & SS-OCT (custom made)                                  & 1310                 \\ \hline
Ossowski et al. \cite{ossowski2015detection} (2015)               & Dynamic         & Statistical properties          & Classification          & Blood                                                                   & SD-OCT (custom made)                                  & 790                  \\ \hline
Roy et al. \cite{roy2015bag} (2015)                               & Static          & Statistical properties          & Classification          & Coronary artery                                                         & SD-OCT (CV-M2, LightLab Imaging Inc)                  & 1320                 \\ \hline
Schmitt et al. \cite{schmitt1999speckle} (1999)                   & Static          & Statistical distributions       & Theoretical modeling    & -                                                                       & -                                                     & -                    \\ \hline
Karamata et al. \cite{karamata2005speckle} (2005)                 & Static          & Statistical distributions       & Theoretical modeling    & -                                                                       & -                                                     & -                    \\ \hline
Mcheik et al. \cite{mcheik2008speckle} (2008)                     & Static          & Statistical distributions       & Segmentation            & Skin                                                                    & SD-OCT (SkinDex 300, ISIS)                            & 1300                 \\ \hline
Kirillin et al. \cite{kirillin2014speckle} (2014)                 & Static          & Statistical distributions       & Theoretical modeling    & Tissue phantoms (polystyrene microspheres)                              & SS-OCT (custom made)                                  & 1310                 \\ \hline
Seevaratnam et al. \cite{seevaratnam2014quantifying} (2014)       & Static          & Statistical distributions       & Classification          & Tissue phantoms (polystyrene microspheres)                              & SS-OCT (Biophotonics and Bioengineering Laboratory's) & 1310                 \\ \hline
Jesus et al. \cite{jesus2015age} (2015)                           & Static          & Statistical distributions       & Classification          & Cornea                                                                  & SD-OCT (Copernicus HR)                                & 850                  \\ \hline
Almasian et al. \cite{almasian2017oct} (2017)                     & Static          & Statistical distributions       & Theoretical modeling    & Tissue phantoms (silica microspheres)                                   & SS-OCT (Santec IVS 2000)                              & 1309                 \\ \hline
Jesus et al. \cite{jesus2017assessment} (2017)                    & Static          & Statistical distributions       & Classification          & Cornea                                                                  & SD-OCT (IOLMaster 700)                                & 850                  \\ \hline
Jesus et al. \cite{jesus2017influence} (2017)                     & Static          & Statistical distributions       & Classification          & Cornea                                                                  & SD-OCT (Copernicus HR)                                & 851                  \\ \hline
Demidov et al. \cite{demidov2019analysis} (2019)                  & Static          & Statistical distributions       & Classification          & Mice (skin)                                                             & SS-OCT (custom made)                                  & 1320                 \\ \hline
Matveev et al. \cite{matveev2019oct} (2019)                       & Static          & Statistical distributions       & Classification          & Mice (skin)                                                             & SS-OCT (custom made)                                  & 1320                 \\ \hline
Matveev et al. \cite{matveev2019assessment} (2019)                & Static          & Statistical distributions       & Classification          & -                                                                       & -                                                     & -                    \\ \hline
Iskander et al. \cite{iskander2020assessing} (2020)               & Static          & Statistical distributions       & Classification          & Cornea                                                                  & SD-OCT (HRT 3, Heidelberg Engineering GmbH)           & 850                  \\ \hline
Cheng et al. \cite{cheng2014histogram} (2014)                     & Dynamic         & Statistical distributions       & Classification          & Phantom: agrose and titanium dioxide / Skin                             & SS-OCT (Thorlabs Inc.)                                & 1300                 \\ \hline
Niemczyk et al. \cite{niemczyk2021effect} (2021)                  & Static          & Statistical distributions       & Classification          & Cornea (porcine eyes)                                                   & SD-OCT (Copernicus REVO)                              & 830                  \\ \hline
Photiou et al. \cite{photiou2017measuring} (2017)                 & Static          & Tissue dispersion               & Classification          & Porcine muscle / Adipose tissues / Colon                                & SS-OCT (custom made)                                  & -                    \\ \hline
Photiou et al. \cite{photiou2017using} (2017)                     & Static          & Tissue dispersion               & Classification          & Porcine muscle / Adipose tissues / Colon                                & SS-OCT (custom made)                                  & 1300                 \\ \hline
Kasaragod et al. \cite{kasaragod2010speckle} (2010)               & Static          & SGLDM                           & Classification          & Tissue phantoms (agar intralipid solution) / Tissue engineered  (skin)  & SS-OCT (custom made)                                  & 1315                 \\ \hline
Gossage et al. \cite{gossage2003texture} (2003)                   & Static          & SGLDM/Frequency Domain methods  & Classification          & Mouse lung                                                              & SS-OCT (custom made)                                  & 1300                 \\ \hline
Gossage et al. \cite{gossage2003texture2} (2003)                  & Static          & SGLDM/Frequency Domain methods  & Classification          & Mouse lung / Bovine tissues                                               & SS-OCT (custom made)                                  & 1300                 \\ \hline
Gossage et al. \cite{gossage2006texture} (2006)                   & Static          & SGLDM/Frequency Domain methods  & Classification          & Tissue phantoms (silica microspheres) / Bovine aorta endothelial cells & SS-OCT (custom made)                                  & 1300                 \\ \hline
Hillman et al. \cite{hillman2006correlation} (2006)               & Static          & Contrast ratio                  & Theoretical modeling    & Tissue phantoms (polystyrene microspheres)                              & SD-OCT (custom made)                                  & 1330                 \\ \hline
Kirkpatrick et al. \cite{kirkpatrick2007quantitative} (2007)      & Dynamic         & Contrast ratio                  & Theoretical modeling / Motion determination    & Engineered tissue                                                       & SD-OCT (custom made)                                  & 843                  \\ \hline
Duncan et al. \cite{duncan2008statistics} (2008)                  & Static          & Contrast ratio                  & Theoretical modeling / Segmentation   & Embryonic chick heart                                                   & -                                                     & -                    \\ \hline
Lee et al. \cite{lee2011speckle} (2011)                           & Static          & Logarithmic intensity contrasts & Theoretical modeling    & Rat liver / Tissue phantoms                                               & SD-OCT (custom made)                                  & 834                  \\ \hline
Motaghiannezam et al. \cite{motaghiannezam2012logarithmic} (2012) & Dynamic         & Logarithmic intensity contrasts & Visualization    & Retina                                                                  & SS-OCT (custom made)                                  & 1060                 \\ \hline
Farhat et al. \cite{farhat2011detecting} (2011)                   & Dynamic         & Speckle correlation             & Motion determination    & Acute myeloid leukemia cells                                            & SS-OCT (Thorlabs Inc.)                                & 1300                 \\ \hline
Liu et al. \cite{liu2013robust} (2013)                            & Dynamic         & Speckle correlation             & Motion determination    & -                                                                       & -                                                     & -                    \\ \hline
Uribe-Patarroyo et al. \cite{uribe2014quantitative} (2014)        & Dynamic         & Speckle correlation             & Motion determination    & Tissue phantoms (intralipid)                                            & SS-OCT (custom made)                                  & 1285                 \\ \hline
De Pretto et al. \cite{de2015new} (2015)                          & Dynamic         & Speckle correlation             & Motion determination    & Milk flow                                                               & SS-OCT (Thorlabs Inc.)                                & 1325                 \\ \hline
Uribe-Patarroyo et al. \cite{uribe2015rotational} (2015)          & Dynamic         & Speckle correlation             & Motion determination    & Endoscopic (esophagus)                                                  & SD-OCT (NvisionVLE)                                   & 1310                 \\ \hline
De Pretto et al. \cite{de2016optical} (2016)                      & Dynamic         & Speckle correlation             & Viscosity determination & Mice blood                                                              & SR-OCT (Thorlabs Inc.)/ SS-OCT (custom made)          & 930  /1325           \\ \hline
Popov et al. \cite{popov2017statistical} (2017)                   & Dynamic         & Speckle correlation             & Viscosity determination & Tissue phantoms                                                         & SD-OCT (custom made)                                  & 1313                 \\ \hline
Ferris et al. \cite{ferris2020forward} (2020)                     & Dynamic         & Speckle correlation             & Motion determination    & Tissue phantoms                                                         & SD-OCT (custom made)                                  & 1290  /1310          \\ \hline

\label{table:articles}
\end{longtable}

\begin{flushleft}

SGLDM = Spatial Gray Level Dependence Matrices;
SD = Spectral Domain;
SS = Swept Source;
SR = Spectral Radar;
NvisionVLE = NvisionVLE Imaging System (NinePoint Medical, Inc., Bedford, MA);
Thorlabs Inc. = Thorlabs Inc. (Newton, NJ);
IOLMaster 700 = IOLMaster 700 (Carl Zeiss Meditec AG, Germany);
CV-M2, LightLab Imaging Inc.= CV-M2, LightLab Imaging Inc. (Westford, MA, USA);
Copernicus HR = Copernicus HR (Optopol, Zawiercie, Poland);
HRT 3, Heidelberg = HRT 3, Heidelberg Engineering GmbH (Heidelberg, Germany);
Copernicus REVO = Copernicus REVO, (Optopol, Zawiercie, Poland).

\end{flushleft}

\end{singlespace}
\end{landscape}
\global\pdfpageattr\expandafter{\the\pdfpageattr/Rotate 0} %
\newpage

\section{Speckle Theory}
\label{sec:speckletheory}

As detailed in the manuscript, one of the main speckle analysis techniques is based on the determination of the speckle pattern intensity probability density function (PDF), since photodetectors measure light intensity and not complex amplitude. A short theoretical introduction on the speckle effect is mandatory to understand why the negative exponential function was historically the first one used to describe the speckle statistics, and to understand their limitations. 

Light speckle is often modelled using the statistical perspective defined by J. W. Goodman \cite{goodman1975statistical}. To deduce the PDFs of the speckle signal amplitude (Gaussian) and its respective intensity (negative exponential) we shall follow Goodman's approach.

Assuming a monochromatic and perfectly polarized light source, for a given temporal instant, we can define the complex amplitude of the electrical field $a$ as \cite{VazThesis}:

\begin{equation} \label{eq:amplitudeSpeckle}
a(x,y,z) = a(x,y,z) e^{i \theta(x,y,z)} ,
\end{equation}

\noindent where a(x,y,z) is the amplitude and $\theta$ the phase. Regardless of the detector spatial position, the amplitude of the electrical field that reaches the detector $A$ corresponds to a sum of $N$ de-phased electrical fields coming from different regions of the tissue:

\begin{equation} \label{eq:result}
    A(x,y,z) =\dfrac{1}{\sqrt{N}} \sum_{k=1}^{N} a_k e^{i \theta_k} .
\end{equation}

\noindent where $a_k$ and $\theta_k, k=1,2,...N$ are the amplitudes and phases forming that field. In order to determine the PDF of the complex amplitude, two assumptions related to the physical mechanisms of speckle are made. First, the amplitude and phase of each phasor are statistically independent of each other. Second, the phases $\theta_k$ are uniformly distributed between $-\pi$ and $\pi$. In terms of physical significance, these assumptions imply that each scattering volume is independent and that the reflection boundary irregularities are larger than the light wavelength.

By splitting the complex amplitude in its real and imaginary parts, it can be shown that both have zero mean and identical variances \cite{goodman1975statistical}. When the number of summed phasors is very large ($N \rightarrow \infty $) the PDF of the real and imaginary part are asymptotically Gaussian as well as their joint PDF:

\begin{equation} \label{eq:jointpdf}
    p(A_{Re},A_{Im}) = p(A_{Re}) \cdot  p(A_{Im}) = \dfrac{1}{\sigma^2 2 \pi} \; e^{-\dfrac{A_{Re}^2+A_{Im}^2}{2\sigma^2}} ,
\end{equation}

\noindent where $A_{Re}$ corresponds to the real part and $A_{Im}$ to the imaginary part of $A(x,y,z)$, and the variance $\sigma^2$ is defined by:

\begin{equation}
    \sigma^2 = \lim_{N \rightarrow \infty} \dfrac{1}{N} \sum_{k=1}^{N}  \dfrac{\langle a_k^2 \rangle}{2} \;.
\end{equation}

The PDF of the light intensity can also be deduced from equation \eqref{eq:jointpdf}. By definition, the light intensity $I$ and the phase $\theta$ are expressed by:

\begin{equation}
I = |A(x,y,z)|^2 = A_{Re}^2 + A_{Im}^2 \; ,
\end{equation}
\begin{equation}
\theta = \tan^{-1} \frac{A_{Im}}{A_{Re}} \; .
\end{equation}

The relation between the intensity PDF and the amplitude PDF can be found by applying random variables transformations:

\begin{equation}
p(I,\theta) = p(A_{Re},A_{Im}) ||J|| = \frac{1}{\sigma^2 4 \pi} \; e^{-\frac{I}{2\sigma^2}} \; ,
\end{equation}

\noindent where $||J||$ is the Jacobian matrix. Recalling the assumptions of independence between intensity and phase, the marginal PDF of the intensity is found using:

\begin{equation}
p(I) = \int_{-\pi}^{\pi} p(I,\theta) d \theta = \frac{1}{2 \sigma^2} \; e^{-\frac{I}{2\sigma^2}} .
\end{equation}

By computing the first and second order moments of the amplitude, the field amplitude variance can be expressed by \cite{VazThesis}:

\begin{equation}
     \sigma^2 = \langle I \rangle / 2 \; ,
\end{equation}

\noindent resulting in a negative exponential PDF which is characteristic of a fully developed speckle pattern in perfect conditions \cite{schmitt1999speckle}:

\begin{equation}
p(I) = \frac{1}{\langle I \rangle} e^{-\frac{I}{\langle I \rangle}} \; .
\end{equation}

\end{appendices}

\end{document}